\documentclass[osajnl,preprint,showpacs,superscriptaddress,12pt]{revtex4-1} 
\usepackage{amsmath,amssymb,graphicx}
\begin{document}

\title{The justification of the `Two-Level Approximation' in strong light-matter interaction}

\author{Faheel Ather Hashmi}\email{Corresponding author: faheel4@comsats.edu.pk}
\affiliation{Beijing Computational Science Research Center, Beijing 100084, P.R. China}
 \affiliation{Department of Physics, COMSATS Institute of Information Technology,  Islamabad, Pakistan}

\author{Shi-Yao Zhu}
\affiliation{Beijing Computational Science Research Center, Beijing 100084, P.R. China}

\begin{abstract}
We investigate the influence of the additional third level on the dynamic evolution of a Two-Level system interacting with a coherent field in the strong coupling regime where Rotating Wave Approximation is not valid. We find that the additional level has great influence on the evolution of the system population. Our results show that the Two-Level model is not a good approximation in this strong light-matter coupling regime. We further investigate the parameter space where the Two-Level model can still be justified. 
\end{abstract}

\maketitle 

\newcommand{\ket}[1]{\left|#1\right\rangle}
\newcommand{\bra}[1]{\left\langle#1\right|}
\newcommand{\abs}[1]{\left|#1\right|}
\newcommand{\op}[2]{\ket{#1}\bra{#2}}

\newcommand{\omgab}{\omega_{ab}}
\newcommand{\omgac}{\omega_{ac}}
\newcommand{\omgz}{\omega_{0}}
\newcommand{\opH}{\hat{H}}
\newcommand{\opHz}{\hat{H_0}}
\newcommand{\opb}{\hat{b}}
\newcommand{\opbdag}{\hat{b}^{\dagger}}
\newcommand{\gab}{g_{ab}}
\newcommand{\gac}{g_{ac}}
\newcommand{\paren}[1]{\left(#1\right)}
\newcommand{\saren}[1]{\left[#1\right]}
\newcommand{\phsabp}{e^{i \Delta_{ab}t}}
\newcommand{\phsabpd}{e^{i \Delta_{ab}^{\prime}t}}
\newcommand{\phsabm}{e^{-i \Delta_{ab}t}}
\newcommand{\phsabmd}{e^{-i \Delta_{ab}^{\prime}t}}

\newcommand{\phsacp}{e^{i \Delta_{ac}t}}
\newcommand{\phsacpd}{e^{i \Delta_{ac}^{\prime}t}}
\newcommand{\phsacm}{e^{-i \Delta_{ac}t}}
\newcommand{\phsacmd}{e^{-i \Delta_{ac}^{\prime}t}}

\section{Introduction}
`Two-level approximation' is a very convenient tool to study light-matter 
interaction because of its simplicity and because of its applicability to a
 large number of real light matter interaction scenarios.
This approximation together with Rotating Wave Approximation (RWA) can be easily
justified when two distinct levels of a system weakly interact with the light
field at resonance or quasi-resonance. 
The other levels of the system are far apart in energy and can be ignored 
along with the `Counter Rotating Terms (CRT)'.  However, RWA becomes 
invalid in strong light-matter coupling regime. This strong 
coupling regime is becoming more accessible for experiments 
\cite{niemczyk10,forn10,you11,you13}, and hence there is a surge of interest 
in going beyond RWA (taking into account the effects of CRT) in the 
description of the interaction \cite{casanova10,albert12}. The breakdown of 
RWA and the role of CRT has also been 
reported \cite{larson12,wang09}. In the strong coupling regime, the two-level 
system without RWA has been very extensively studied 
\cite{irish07,larson07,werlang08,Braak11,zubairy88,Milonni83,Spreeuw90}. 
In these studies the counter-rotating terms are taken into account, 
but the two-level approximation is still adopted.
 The counter-rotating terms
 involve the contribution proportional to $1/(\omega_{ab}+\omega_0)$
where $\omega_{ab}$ is the transition frequency of the system  and $\omega_0$ is the frequency of the field. 
If there is an additional level of the system (say $\ket c$
with transition frequency $\omega_{ac}$), the rotating wave terms on this transition will result
in the contribution proportional to $1/(\omega_{ac}-\omega_0)$, which can be
larger than the contribution proportional to  $1/(\omega_{ab}+\omega_0)$.
Consequently, neglecting the additional levels is questionable.  
Therefore, in the strong coupling regime where 
CRT effects have to be taken into account, the question `whether  the effects due to additional levels can be neglected?' needs to be treated carefully.
 In this paper, we consider the 
effects of the third level on the population dynamics of a two-level 
system in strong coupling regime, and focus on when the third level can be  neglected.

The three level system interacting with resonant or quasi-resonant fields beyond RWA has also been 
studied, mainly in the context of trapping dark state in $\Lambda$ configuration \cite{unanyan00,ho85,matisov95,xiaohong12,sanchez04}.
However, these studies concerned with the effects of CRT terms on the three level system, 
and did not discuss the consequence of the third level on the evolution.
In the present work we consider the three level `V-system', where the third 
level is non-resonant and is far away in energy from the two-level transition,
and study the effects of the third level on the population dynamics.
Our focus is on when the effects of the third level will diminish in the 
strong coupling regime. 

\section{Model}
\begin{figure}[htbp]
\begin{center}
\includegraphics{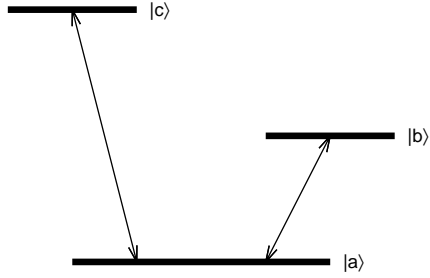}
\end{center}
\caption{A three level system in `V' configuration. We are interested in the effects of the additional level $\ket c$ on the population dynamics of level $\ket b$.}
\label{fig:1}
\end{figure}
Consider a three level system in `V' configuration as shown in FIG. \ref{fig:1} interacting with a single mode quantized field. The Hamiltonian of the system can be written as (with $\hbar=1$)
\begin{equation}\label{eqHamil}
\begin{aligned}
\opH =& \omgab \op bb + \omgac \op cc + \omgz \opbdag\opb 
\\	& +\gab \paren{ \opbdag + \opb } \saren{\op ab + \op ba }
\\	& +\gac \paren{ \opbdag + \opb } \saren{\op ac + \op ca },
\end{aligned}
\end{equation}
where $\omgab$ and $\omgac$ are the transition frequencies for the excited states and $\opbdag$, $\opb$, and $\omgz$ are the creation and annihilation operators and the frequency for the field. The coupling constants $\gab$ and $\gac$ are real. We choose $\ket {n,l}$  as the basis to write the system ket where $n$ is the number of photon and $l=\{a,b,c\}$ denotes the level of the system. 
The system ket can be written as
\begin{align}
\ket{\Psi}= \sum_n^{\infty}\sum_{p=0}^{1} e^{-i\opHz t} a_{2n+p} \ket{2n+p,a} + b_{2n+p}\ket{2n+p,b} + c_{2n+p}\ket{2n+p,c}
\label{eqKet}
\end{align}
In this expression $\opHz=\omgab \op bb + \omgac \op cc + \omgz \opbdag\opb $ is the non-interacting part of the Hamiltonian in eq(\ref{eqHamil}). The system ket eq(\ref{eqKet}) actually consists of two independent kets that separately satisfy the Schr\"{o}dinger equation. This is because of the fact that the Hamiltonian admits even and odd parity chains of evolution like the JCM Hamiltonian with CRT \cite{casanova10}. One such chain consists of the states with even number of photons in $\ket a$ and odd number of photons in $\ket b$ and $\ket c$, and the other chain has states with odd number of photons in $\ket a$ and even number of photons in $\ket b$ and $\ket c$. The time evolution of the amplitudes is given by
\begin{equation}
\label{eqMaster}
\begin{aligned}
i \dot{a}_{2n+p} &= \gab \left[\sqrt{2n+p} \, b_{2n+p-1} \phsabm + \sqrt{2n+p+1} \,b_{2n+p+1}\phsabmd \right]
\\		&+ \gac \left[\sqrt{2n+p} \, c_{2n+p-1} \phsacm + \sqrt{2n+p+1} \,c_{2n+p+1}\phsacmd \right]
\\
i \dot{b}_{2n+p} &= \gab \left[\sqrt{2n+p+1} \,a_{2n+p+1} \phsabp + \sqrt{2n+p} \,a_{2n+p-1}\phsabpd\right]
\\
i \dot{c}_{2n+p} &= \gac \left[\sqrt{2n+p+1} \,a_{2n+p+1} \phsacp + \sqrt{2n+p} \,a_{2n+p-1}\phsacpd\right]
\end{aligned}
\end{equation}
Here $\Delta_{ab}= \omega_{ab}-\omega_0$, $\Delta_{ab}^{\prime}=\omega_{ab}+\omega_0$, $\Delta_{ac}= \omega_{ac}-\omega_0$, and $\Delta_{ac}^{\prime}=\omega_{ac}+\omega_0$. $p$ in each equation is either $0$ for state $\ket a$ and $1$ for the states $\{\ket b, \ket c \}$, or $1$ for the state $\ket a$ and $0$ for excited states. In the following we solve these coupled differential equations numerically for a system that is initially in state $\ket b$, and for the field that is in a coherent state with average photon number $\bar n=10$. The population in state $\ket b$ $=\sum_n \abs{b_n}^2$ is calculated in the presence ($\gac\ne 0$) and the absence ($\gac=0$) of the additional level $\ket c$.  The summation in eq (\ref{eqKet})  is truncated at $n=200$. 

\section{Results \& Discussion}
\begin{figure}[htbp]
\begin{center}
\includegraphics[width=0.4\textwidth]{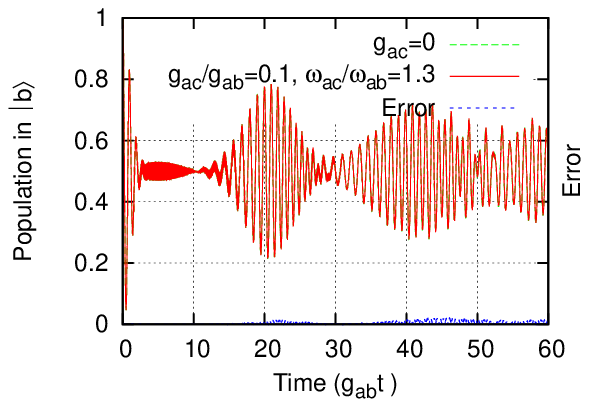}
\includegraphics[width=0.4\textwidth]{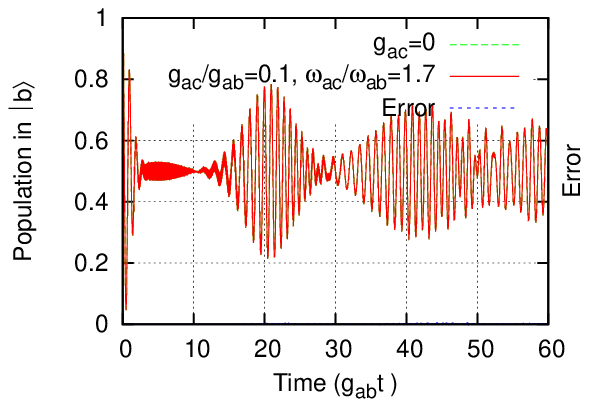}
\\
\includegraphics[width=0.4\textwidth]{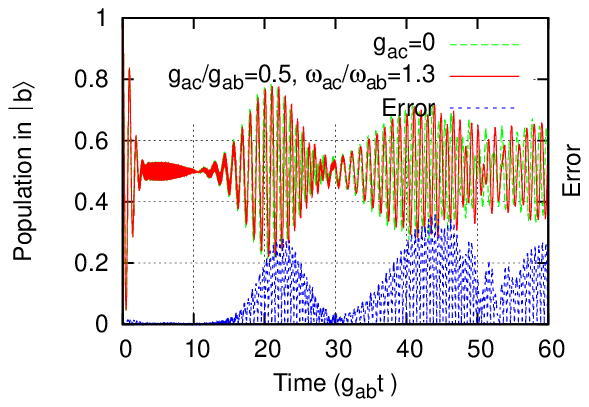}
\includegraphics[width=0.4\textwidth]{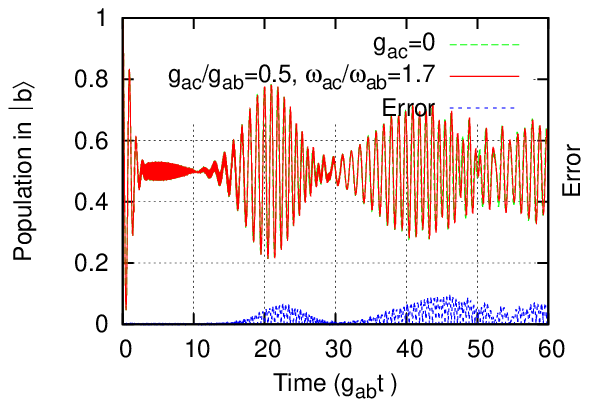}
\\
\includegraphics[width=0.4\textwidth]{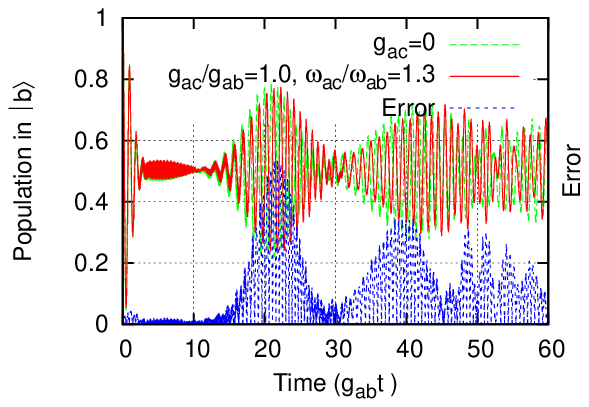}
\includegraphics[width=0.4\textwidth]{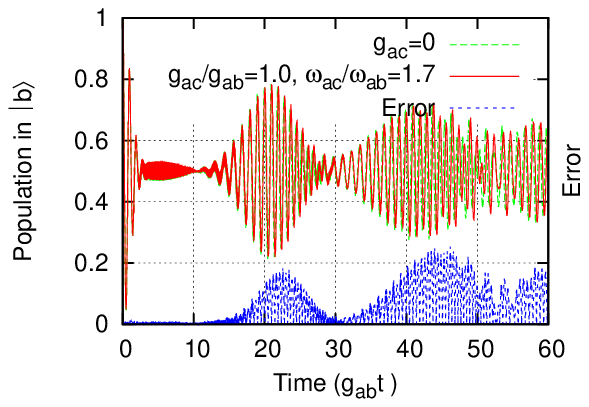}
\end{center}
\caption{(color online) The population dynamics of state $\ket b$ in the absence (dashed-green) and presence (solid-red)  of the additional level $\ket c$. The difference of the two (dotted-blue) is also shown. The parameters are $\gab=0.02 \omega_{ab}$, $\omega_0=\omega_{ab}$, and $\bar n=10$.}
\label{fig:2}
\end{figure}
In FIG. \ref{fig:2} we show the population dynamics in the state $\ket b$
  in the absence (dashed-green), and presence (solid-red) of the additional level $\ket c$ for a moderately strong coupling $\gab=0.02\omega_{ab}$.
The difference of the two signifies the importance of additional level effects, and is shown in dotted (blue) curve.
The dashed (green) curves exhibit traditional collapse and revival of the 
dynamics in the two-level model 
\cite{eberly80} superimposed by strong oscillations due to counter-rotating terms \cite{zubairy88}.
This behavior is modified by the additional level $\ket c$. 
With the additional level present, and for short times, the dynamics follow the two-level case but at longer times
strong differences appear in the dynamics. These differences can give us the criteria for the justification of
`Two-Level Approximation' in the strong coupling regime. As we can see in the figure that for weak additional level coupling 
$\gac/\gab\rightarrow 0$,  the differences are small, and hence the approximation can be justified.
However, for $\gab/\gac=0.5$, a level placed at $\omgac/\omgab=1.3$ strongly modifies the dynamics, and raises 
concerns on the two-level model. For still higher coupling $\gac=\gab$, even a level placed at $\omgac/\omgab=1.7$ can not be ignored.

\begin{figure}[htbp]
\begin{center}
\includegraphics[width=0.5\textwidth]{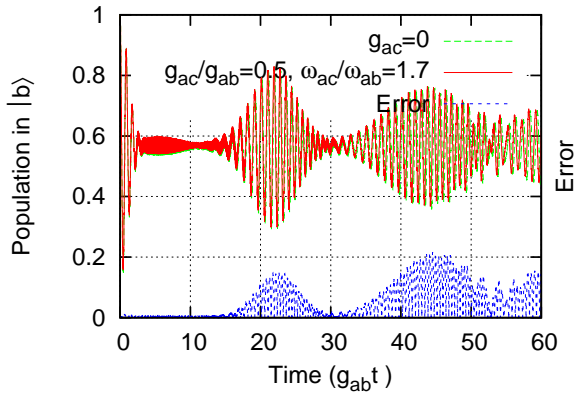}
\end{center}
\caption{(color online) Additional level effects when the two-level transition $\ket a \leftrightarrow \ket b$ is detuned from resonance.
Parameters are $\gab=0.02 \omgab$, $\omgz= 0.95\omgab$, and $\bar n=10$} 
\label{fig:3}
\end{figure}

The additional level effects become more prominent when the two-level system is detuned from the resonance as shown in FIG. \ref{fig:3}.
 Here we make the field non-resonant on $\ket a\leftrightarrow \ket b$ transition by taking $\omgz=0.95\omgab$. The resulting difference
in dynamics are more pronounced than the corresponding case in FIG. \ref{fig:2} (with parameters $\gac/\gab=0.5$ and $\omgac/\omgab=1.7$).

\begin{figure}[htbp]
\begin{center}
\centerline{\includegraphics[width=.8\columnwidth]{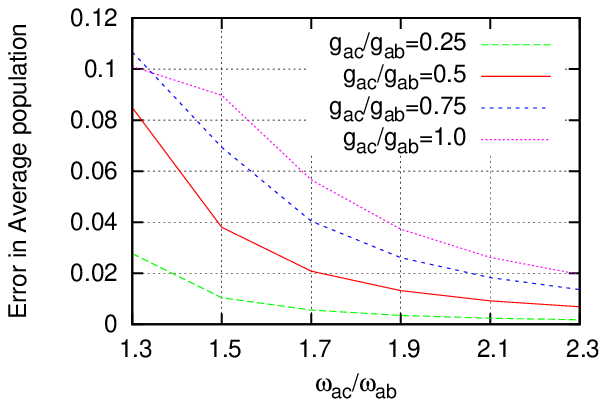}}
\end{center}
\caption{(color online) The average of the absolute population difference in state $\ket b$ in the presence and absence of the additional level.
The parameters are $\gab=0.02 \omega_{ab}$, and $\bar n=10$.}
\label{fig:5}
\end{figure}

Finally, in FIG. \ref{fig:5} we plot an estimate of the error in population due to the additional level. This error is calculated  by taking 
the average of the absolute difference in population dynamics in 
the presence and absence of the additional level, and is plotted as 
the function of the placement of the additional level $\omgac/\omgab$ for various coupling strengths $\gac/\gab$ with $\gab=0.02\omgab$.
Here, we again see that the error due to the additional level can be significant even if it is placed as far as $\omgac/\omgab=1.3$ for non-vanishing 
coupling $\gac$.
 However, the error decreases as the additional level goes further away
 in energy, and as the coupling of the ground state to this additional
 level $\gac$ decreases in comparison with $\gab$.

\section{Conclusion}
We have shown that the additional level can have important effects on the 
population dynamics of the two-level model in the strong coupling regime.
However, the `Two-Level approximation' can still be valid when the additional level is far away (few units) from resonance, and the coupling with the additional level is weaker compared to the coupling with the two-level model. 
The effects of the additional level may become very crucial in
experiments in the strong coupling regime where precise control of the quantum dynamics is required. Our results suggest that in such strong light-matter 
interaction  regime, the `Two-Level approximation' should be used with care. 
\newpage



\begin{thebibliography}{21}
\newcommand{\enquote}[1]{``#1''}
\bibitem{niemczyk10}
T.~Niemczyk, F.~Deppe, H.~Huebl, E.~P. Menzel, F.~Hocke, M.~J. Schwarz, J.~J.
  Garcia-Ripoll, D.~Zueco, T.~H\"{u}mmer, E.~Solano, A.~Marx, and R.~Gross,
  \enquote{Circuit quantum electrodynamics in the ultrastrong-coupling regime,}
  Nat Phys \textbf{6}, 772--776 (2010).

\bibitem{forn10}
P.~Forn-D\'{i}az, J.~Lisenfeld, D.~Marcos, J.~Garc\'{i}a-Ripoll, E.~Solano,
  C.~Harmans, and J.~Mooij, \enquote{Observation of the bloch-siegert shift in
  a qubit-oscillator system in the ultrastrong coupling regime,} Phys. Rev.
  Lett. \textbf{105}, 237001 (2010).

\bibitem{you11}
J.~Q. You and F.~Nori, \enquote{Atomic physics and quantum optics using
  superconducting circuits,} Nature \textbf{474}, 589--597 (2011).

\bibitem{you13}
Z.-L. Xiang, S.~Ashhab, J.~Q. You, and F.~Nori, \enquote{Hybrid quantum
  circuits: Superconducting circuits interacting with other quantum systems,}
  Rev. Mod. Phys. \textbf{85}, 623--653 (2013).

\bibitem{casanova10}
J.~Casanova, G.~Romero, I.~Lizuain, J.~Garc\'{i}a-Ripoll, and E.~Solano,
  \enquote{Deep strong coupling regime of the jaynes-cummings model,} Phys.
  Rev. Lett. \textbf{105}, 263603 (2010).

\bibitem{albert12}
V.~Albert, \enquote{Quantum rabi model for n-state atoms,} Phys. Rev. Lett.
  \textbf{108}, 180401 (2012).

\bibitem{larson12}
J.~Larson, \enquote{Absence of vacuum induced berry phases without the rotating
  wave approximation in cavity {QED},} Phys. Rev. Lett. \textbf{108}, 033601
  (2012).

\bibitem{wang09}
D.-W. Wang, L.-G. Wang, Z.-H. Li, and S.-Y. Zhu, \enquote{Anti-zeno-effect
  recovery and lamb-shift modification in modified vacuum,} Phys. Rev. A
  \textbf{80}, 042101 (2009).

\bibitem{irish07}
E.~Irish, \enquote{Generalized rotating-wave approximation for arbitrarily
  large coupling,} Phys. Rev. Lett. \textbf{99}, 173601 (2007).

\bibitem{larson07}
J.~Larson, \enquote{Dynamics of the {Jaynes–Cummings} and rabi models: old
  wine in new bottles,} Phys. Scr. \textbf{76}, 146 (2007).

\bibitem{werlang08}
T.~Werlang, A.~V. Dodonov, E.~I. Duzzioni, and C.~J. Villas-Bôas,
  \enquote{Rabi model beyond the rotating-wave approximation: Generation of
  photons from vacuum through decoherence,} Phys. Rev. A \textbf{78}, 053805
  (2008).

\bibitem{Braak11}
D.~Braak, \enquote{Integrability of the rabi model,} Phys. Rev. Lett.
  \textbf{107}, 100401 (2011).

\bibitem{zubairy88}
K.~Zaheer and M.~Zubairy, \enquote{Atom-field interaction without the
  rotating-wave approximation: A path-integral approach,} Phys. Rev. A
  \textbf{37}, 1628--1633 (1988).

\bibitem{Milonni83}
P.~Milonni, J.~Ackerhalt, and H.~Galbraith, \enquote{Chaos in the semiclassical
  n-atom jaynes-cummings model: Failure of the rotating-wave approximation,}
  Phys. Rev. Lett. \textbf{50}, 966--969 (1983).

\bibitem{Spreeuw90}
R.~J.~C. Spreeuw, N.~J. van Druten, M.~W. Beijersbergen, E.~R. Eliel, and J.~P.
  Woerdman, \enquote{Classical realization of a strongly driven two-level
  system,} Phys. Rev. Lett. \textbf{65}, 2642--2645 (1990).

\bibitem{unanyan00}
R.~Unanyan, S.~Gu\'{e}rin, and H.~Jauslin, \enquote{Coherent population
  trapping under bichromatic fields,} Phys. Rev. A \textbf{62}, 043407 (2000).

\bibitem{ho85}
T.-S. Ho and S.-I. Chu, \enquote{Semiclassical many-mode floquet theory. {III.}
  {SU(3)} dynamical evolution of three-level systems in intense bichromatic
  fields,} Phys. Rev. A \textbf{31}, 659--676 (1985).

\bibitem{matisov95}
B.~Matisov, I.~Mazets, and L.~Windholz, \enquote{Coherent population trapping
  beyond the rotating-wave approximation,} Quantum Semiclass. Opt. \textbf{7},
  449 (1995).

\bibitem{xiaohong12}
X.~L. Xiaohong~Li, T.~L. Tao~Liu, and K.~W. Kelin~Wang, \enquote{Discussions of
  ground state in lambda-type three-level system,} Chin. Opt. Lett.
  \textbf{10}, S12701--312704 (2012).

\bibitem{sanchez04}
B.~Sanchez and T.~Brandes, \enquote{Matrix perturbation theory for driven
  three-level systems with damping,} Ann. Phys. \textbf{13}, 569--594 (2004).

\bibitem{eberly80}
J.~Eberly, N.~Narozhny, and J.~Sanchez-Mondragon, \enquote{Periodic spontaneous
  collapse and revival in a simple quantum model,} Phys. Rev. Lett.
  \textbf{44}, 1323--1326 (1980).

\end{thebibliography}

\end{document}